\documentclass[a4paper,superscriptaddress,twocolumn,pra,showpacs,floatfix]{revtex4}
\usepackage{epsfig}
\usepackage{graphicx}
\usepackage{epstopdf}
\usepackage{amsmath} 
\usepackage{bm}
\usepackage{color}

\begin{document}
\title{Engineering interband transport by time-dependent disorder}

\author{Ghazal Tayebirad}
\affiliation{Institut f\"ur Theoretische Physik, Universit\"at Heidelberg, Philosophenweg 19, 69120 Heidelberg, Germany}

\author{Riccardo Mannella}
\affiliation{Dipartimento di Fisica `E. Fermi', Universit\`{a} di Pisa, Largo Pontecorvo 3, 56127 Pisa, Italy} 

\author{Sandro Wimberger}
\affiliation{Institut f\"ur Theoretische Physik, Universit\"at Heidelberg, Philosophenweg 19, 69120 Heidelberg, Germany}
\affiliation{Heidelberg Center for Quantum Dynamics, Philosophenweg 12, 69120 Heidelberg, Germany}

\date{\today}

\begin{abstract}
We show how the evolution of atoms in a tilted lattice can be changed and controlled by phase noise on the lattice. Dependent on the characteristic parameters of the noise, the interband transport can either be suppressed or enhanced, which is of interest for high precision control in experimental realization with Bose-Einstein condensates. The effect of the noise on the survival probability in the ground band is summarized in a scaling plot stressing the universality of our results.
\end{abstract}


\maketitle

Ultracold atoms in optical lattices have opened a possibility to investigate effects which were not previously observable on ordinary matter crystals, at least not in such a controlled way, like Bloch oscillations, Wannier-Stark ladders, and Landau--Zener tunneling \cite{Dahan96,Raizen97,Morsch06,Sias07,Gustavsson08,Zenesini09,Naegerl10}. Loading Bose--Einstein condensates (BEC) into optical lattices provides an optimal control of the studied systems by optical means. More complicated potentials can be realized by adding further lattice beams \cite{Santos04,Salger07}. By superimposing laser beams from different directions and with different frequencies, it is possible to generate various lattice geometries \cite{Morsch06,Salger07,Bloch07,Roati08,Inguscio10}. The question arises how to control the dynamics of particles by quasi-periodic potentials (possibly time-dependent or even stochastic ones). From the theoretical point of view, a variety of phenomena is expected to occur in these systems, such as Anderson localization~\cite{Billy08,Roati08} and the quantum transition to the Bose glass phase originating from the interplay of interaction and spatial, but static disorder~\cite{Scalettar91,Inguscio10}.

In this Rapid Communication, we investigate the Landau--Zener tunneling of a BEC from the ground band of a lattice into higher bands and, finally, into the energetic continuum. We show that control on the interband coupling is not only achieved by varying the lattice parameters, the initial conditions \cite{Sias07,Zenesini09}, or by stimulated transitions \cite{Bloch07}, but also by additional noise which produces a time-dependent disordered potential \cite{Tayebirad10-2}. More specifically, we study the time evolution of the survival probability of a BEC loaded into a quasi-1D geometry built of a stochastic potential and a static Stark force. While preliminary results reported in \cite{Tayebirad10-2} imply that exponentially correlated noise is not efficient in controlling the interband decay, we present here a universal prediction based on harmonic noise with a well-defined frequency range. By adapting the noise frequency to the system's scales we demonstrate that stochastic noise can be used to engineer the decay very efficiently. % without changing system intrinsic parameters.

We focus on the dynamics of a sufficiently dilute non-interacting BEC. In the absence of atom-atom interactions, the temporal evolution of a BEC in a tilted one-dimensional lattice is described by the single-particle Hamiltonian 
\begin{equation}
\label{1DHamil}
H=-\frac{\hbar^{2}}{2M}\frac{d^2}{dx^2}+V_{\rm s}\left(x,t\right)+Fx  \,,
\end{equation}
where $M$ is the atomic mass and $F$ the Stark force. $V_{\rm s}\left(x,t\right)$ is a stochastic potential given by the following time-dependent bichromatic lattice 
\begin{equation}
V_{\rm s}\left(x,t\right)=\alpha V\left\lbrace\sin^{2}\left(\frac{\pi x}{d_{\rm L}}\right)+\sin^{2}\left(\frac{\pi x}{d^{'}_{\rm L}}+\phi(t)\right)\right\rbrace  \,.
\label{DisorderLattice}
\end{equation}
It consists of two spatially periodic potentials with incommensurate spacings, $d_{\rm L}$ for the ``reference'' lattice and $d^{'}_{\rm L}=d_{\rm L}(\sqrt{5}-1)/2$ for the secondary lattice. The time-dependence arises from the time-dependent stochastic phase $\phi(t)$. For constant $\phi$, this is the system with a quasi-disordered lattice which was realized by Roati~\textit{et al.}~\cite{Roati08}. The necessary phase control is easily achieved in experiments for the time scales studied here (about $10 \rm \mu s \ldots 0.1 \rm s$) using modulated lattice beams \cite{phase}. The depths of the two lattices are considered to be comparable and for convenience equal amplitudes $\alpha V$ are chosen, where $V$ is the lattice depth of the reference lattice and $\alpha$ is a renormalization factor (which will be defined below). The recoil energy $E_{\rm{rec}}= p_{\rm rec}^2/2M$, with $p_{\rm rec} = \pi\hbar/d_{\rm L}$, is the characteristic energy scale of the optical reference lattice \cite{Morsch06}, and $V_{0}=V/E_{\rm rec}$ and $F_{\rm 0}=Fd_{\rm L}/E_{\rm rec}$ are dimensionless quantities in this energy unit. We assume a typical experimental situation \cite{Morsch06,Sias07,Zenesini09}, for which the initial state is a condensate's wave function relaxed in a harmonic trap and then loaded adiabatically into the lattice given by $V_{\rm s}(x,t=0)$. The trap is shallow such that many lattice sites are initially populated. It is turned off for the further time evolution.

For the original Wannier-Stark problem including just the first term on the r.h.s. of Eq.~\eqref{DisorderLattice}, the wave packet prepared in the ground band oscillates periodically in time with the Bloch period $T_{\rm B}=2\pi\hbar/(F_0E_{\rm rec})$ and frequency $\omega_{\rm B}=2\pi/T_{\rm B}$~\cite{Glueck02}. However, for sufficiently large $F$, the BEC undergoes Landau--Zener tunneling  after each Bloch period at the band edge, where the band gap has its smallest value \cite{Morsch06,Wimberger05}. Such a phenomenon gives rise to a step-like structure in the time-resolved survival probability $P_{\rm sur}(t)$ of the atoms in the ground band \cite{Zenesini09}. Following \cite{Zenesini09,Sias07,Wimberger05}, we calculate $P_{\rm sur}(t)$ in momentum space by projecting onto the support of the initial state: $P_{\rm sur}(t)= \int^{\infty}_{-p_{\rm c}} dp_{\rm x}\vert\psi(p_{\rm x},t)\vert^{2}$. The upper bound can be safely extended to infinity since the tunneled wave packet moves into the direction of negative momenta \cite{Wimberger05}. The lower bound is given by the parameter $p_{\rm c}$, which can be adapted to stabilize the results with respect to temporal fluctuations. While the true decay starts after a traversal of half the Brillouin zone (the Brillouin zone has a width of $2p_{\rm rec}$), we choose either $p_{\rm c}=3 p_{\rm rec}$ or $p_{\rm c}=5 p_{\rm rec}$, which effectively means that we measure the decay with a delay of $T_{\rm B}$ or $2T_{\rm B}$ respectively. As long as the mean momentum evolves linearly in time with a variance significantly smaller than the width of the Brillouin zone, we can resort to the acceleration theorem~\cite{Dahan96} to relate both ways of measuring. In the following, we introduce the type of stochastic process we use to generate $\phi (t)$ and show universal properties of the survival probability as a function of rescaled parameters.

\begin{figure}%[H]
\begin{center}
   \includegraphics[width=0.9\linewidth,angle=0]{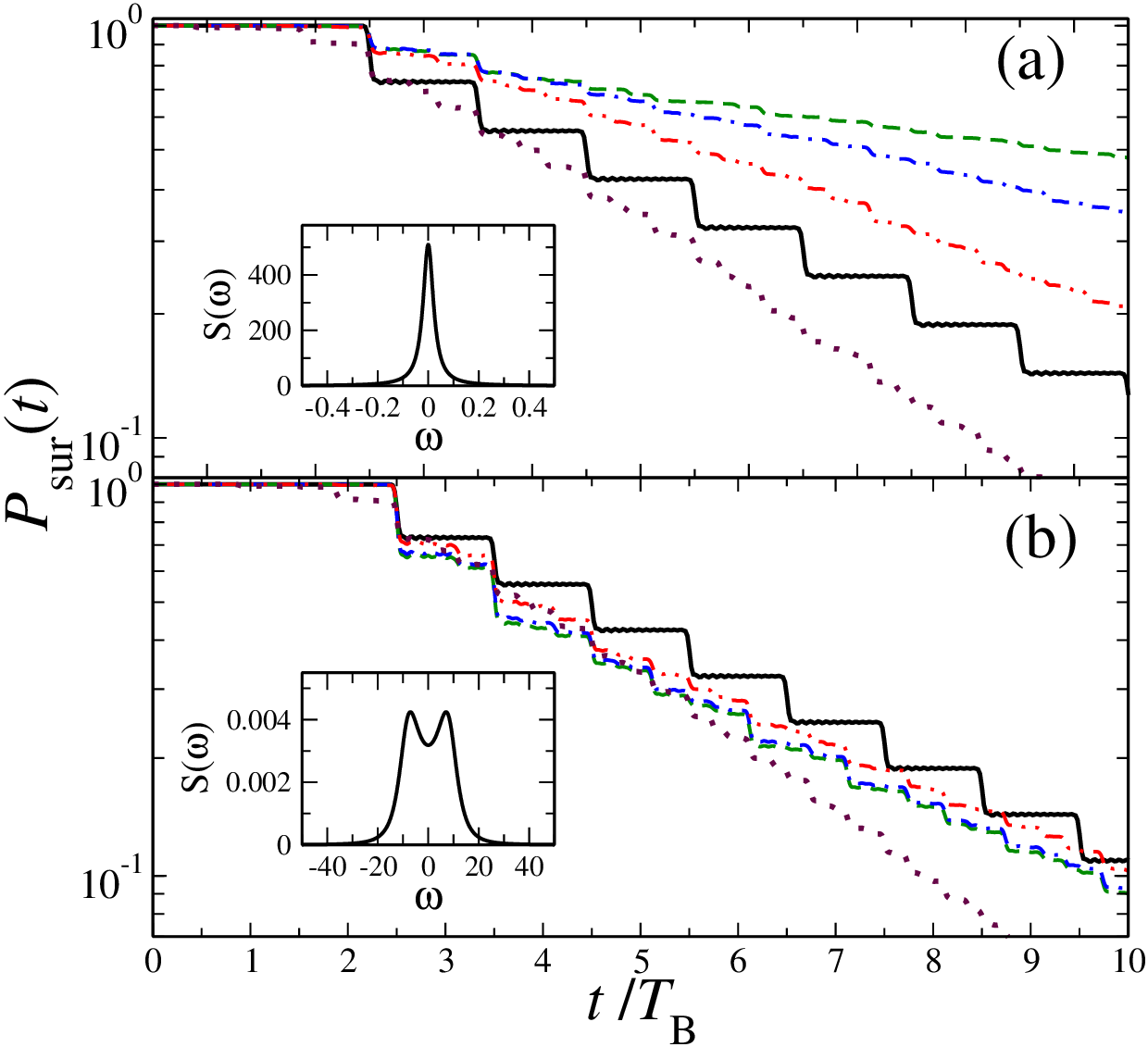}
   \caption{
   (color online) 
$P_{\rm sur}(t)$ $V_{\rm 0}=2.5$, $F_{0}=1.5$ in the presence of harmonic noise with $\Gamma=5/T_{\rm B}$, and $\omega_{0}=0.1/T_{\rm B}$ ((a) -- non-universal regime, c.f. Fig.~\ref{P_surt0_omega_variace} at low values on the $x$ axis) or $\omega_{0}=10/T_{\rm B}$ (b) and for $T=0.01/T_{\rm B}^2$ (green dashed lines), $T=1/T_{\rm B}^2$ (blue dot-dashed lines), $T=10/T_{\rm B}^2$ (red dot-dot-dashed lines), and $T=100/T_{\rm B}^2$ (dotted lines). The result for the reference system is shown by the thick solid lines. The corresponding noise spectra $S(\omega)$ are given in the insets for the case of $T=10/T_{\rm B}^2$.
   }
   \label{HN-POT}
\end{center}
\end{figure}

In general, colored noise can be defined by a suitable spectral distribution. The simplest type can be produced by a harmonic oscillator of angular frequency $\omega_{0}$ and damping rate $\Gamma$, driven by Gaussian white noise. The result is harmonic noise~\cite{Schimansky90}, which is represented by a two dimensional Gauss-Markov process taking the form of the following stochastic differential equations for $\phi(t)$:
\begin{subequations}
\begin{align}
    \dot\phi &= \nu \label{HN_SDE_a} \\    
    \dot\nu  &=  - 2 \Gamma \nu - \omega^{2}_{0} \phi + \xi(t)  \, .
\label{HN_SDE_b}
\end{align}
\label{HN_SDE}
\end{subequations}
\noindent $\xi(t)$ is Gaussian white noise with $\langle\xi(t)\rangle=0$, $\langle\xi(t)\xi(s)\rangle=4\Gamma T\delta(t-s)$, and $T$ measures the strength of the noise. %, and in the limit of white noise, through the fluctuation dissipation theorem, it is related to temperature. 
$\Gamma$ and $\omega_{0}$ have the units of $T_{\rm B}^{-1}$ and $\omega_{\rm B}$ respectively. For notational reasons, we express both of them in the unit of $T_{\rm B}^{-1}$. The variance of the noise is given by $\langle\phi^2(t)\rangle=T/\omega_0^2$, where $\langle\cdot\rangle$ is a time average over a sufficiently large interval (in our simulations $\approx 20\,T_{\rm B}$). The spectral distribution for the harmonic noise is
   \begin{equation}
     S\left(\omega\right) = \frac{2\Gamma T}{\pi(4\Gamma^{2}\omega^{2}+(\omega^{2}-\omega^{2}_{\rm 0})^{2})} \, .  
   \label{spectrum_HN} 
   \end{equation} 
This distribution incorporates two important regimes: ${\rm (i)}$ slowly-varying noise when the oscillation frequency is much less than the damping rate, i.e., $2\Gamma^{2} \gg \omega_{\rm 0}^{2}$, and ${\rm (ii)}$ fast noise, when $2\Gamma^{2} \ll \omega_{\rm 0}^{2}$. The spectral distributions corresponding to these limits are shown in the insets of Fig.~\ref{HN-POT}. In the regime ${\rm (i)}$, the noise recovers the exponentially correlated noise~\cite{Schimansky90,Ornstein-Uhlenbeck} with a Lorentzian-like distribution peaked around zero frequency (see Fig.~\ref{HN-POT}(a)). The spectral distribution in the regime of fast noise (see Fig.~\ref{HN-POT}(b)) peaks at $\tilde{\omega}_0=\sqrt{\omega^{2}_{\rm 0}-2\Gamma^2}$ with a finite width $\Delta\tilde{\omega}_{0}\approx 2\sqrt{\Gamma\sqrt{\omega_0^2-\Gamma^2}}$. To control the evolution in the potential of Eq.~\eqref{DisorderLattice} the most relevant noise realization is this latter regime (ii), for which the above mentioned energy scales can be matched with the scales of the original Wannier-Stark system.

For very fast noise, the particle effectively averages over the time-dependent potential. Assuming a Gaussian distribution of the fast varying phase with variance $\langle\phi^2 \rangle$ and integrating over the phase as random variable, an effective potential is calculated as
\begin{equation}
 V_{\rm eff}\left(x\right)=\alpha V\left\lbrace\sin^{2}\left(\frac{\pi x}{d_{\rm L}}\right)+\beta \sin^{2}\left(\frac{\pi x}{d^{'}_{\rm L}}\right)\right\rbrace \, , 
 \label{eq:eff} 
\end{equation} 
where $\beta=\exp\left(-2 \langle\phi^2 \rangle\right)$. The effect of fast noise is hence to renormalize the amplitude of the second lattice by a factor $\beta$. The parameter $\alpha$ is introduced to compare better the dynamics in the potential given by Eq.~\eqref{DisorderLattice} with the dynamics in the reference system, in which just $V\left(x\right)=V\sin^{2}\left(\pi x/d_{\rm L}\right)$ is present. For this, $\alpha$ is chosen such that the spatial second moments of $V(x)$ and $V_{\rm eff}(x)$ are equal \cite{Tayebirad10-2}.

We study the temporal behavior of the survival probability $P_{\rm sur}(t)$, which is the main experimental observable for our system \cite{Sias07,Zenesini09,Wimberger05}, in a broad range of noise parameters. We use a reference system with typical experimental values of $V_{0}=2.5$ (giving an average band gap $\Delta E\approx2.5\,E_{\rm rec}$) and $F_{0}= 1.5$  \cite{Sias07,Zenesini09}. $\omega_{0}$ and $\Gamma$, are chosen in the range of $0.01/T_{\rm B}$ and $300/T_{\rm B}$, which covers both regimes of fast and slow noise. The stochastic nature makes it necessary to average the results over a sufficient number of statistical realizations. For the parameters investigated here, 20 realizations for fixed noise parameters turned out to well stabilize statistical fluctuations. The lattice parameters $V_{\rm0}$ and $F_{\rm0}$ can be adapted to fulfill the condition of resonantly enhanced tunneling (RET) in the reference system, i.e., $F_{\rm 0}\approx n\Delta E/E_{\rm rec}$, with an integer $n$~\cite{Glueck02,Sias07,Wimberger05}, or away from this special condition (non-RET). Since at RET the transport from the ground band to higher bands is already enhanced by energetic quasi degeneracies~\cite{Glueck02}, and the noise is likely to drive the system out of RET conditions, the decay is expected to degrade for typical parameters in this case \cite{Tayebirad10-2}. For our non-RET conditions $V_{0}=2.5$ and $F_{0}= 1.5$, it is shown in the following that a faster decay can be easily induced by harmonic noise, since there is no competition between the two effects (enhancement by RET and noise).

The stochastic potential $V_{\rm s}\left(x,t\right)$ is determined by the spectral distributions $S(\omega)$ of the noise, which are depicted in the insets of 
Fig.~\ref{HN-POT}(a) and (b) for the two different regimes of noise, respectively. For a small oscillation frequency $\omega_0=0.1/T_{\rm B}$ compared with the damping rate $\Gamma=5/T_{\rm B}$, $S(\omega)$ has a very narrow peak at zero frequency. The phase itself is slowly varying with time in case (a), whilst it shows faster fluctuations in case (b). The effect of the noise on the temporal evolution of $P_{\rm sur}(t)$ is compared in Figs.~\ref{HN-POT}(a) and (b) to the one for the noise-free reference system with its characteristic step-like structure (solid lines) \cite{Wimberger05,Zenesini09}. As seen, the harmonic noise tends to wash out the step structure after a few Bloch oscillations. Moreover, it leads to a systematic enhancement of the tunneling rate for the largest noise amplitude $T$ in both cases (a,b). In the regime of fast noise (Fig.~\ref{HN-POT}(b)) the tunneling rate is always enhanced with respect to the reference system, and here it is essentially independent of $T$ (see also Fig.~\ref{P_surt0_omega_T} below).

\begin{figure}
\begin{center}
   \includegraphics[width=0.9\linewidth,angle=0]{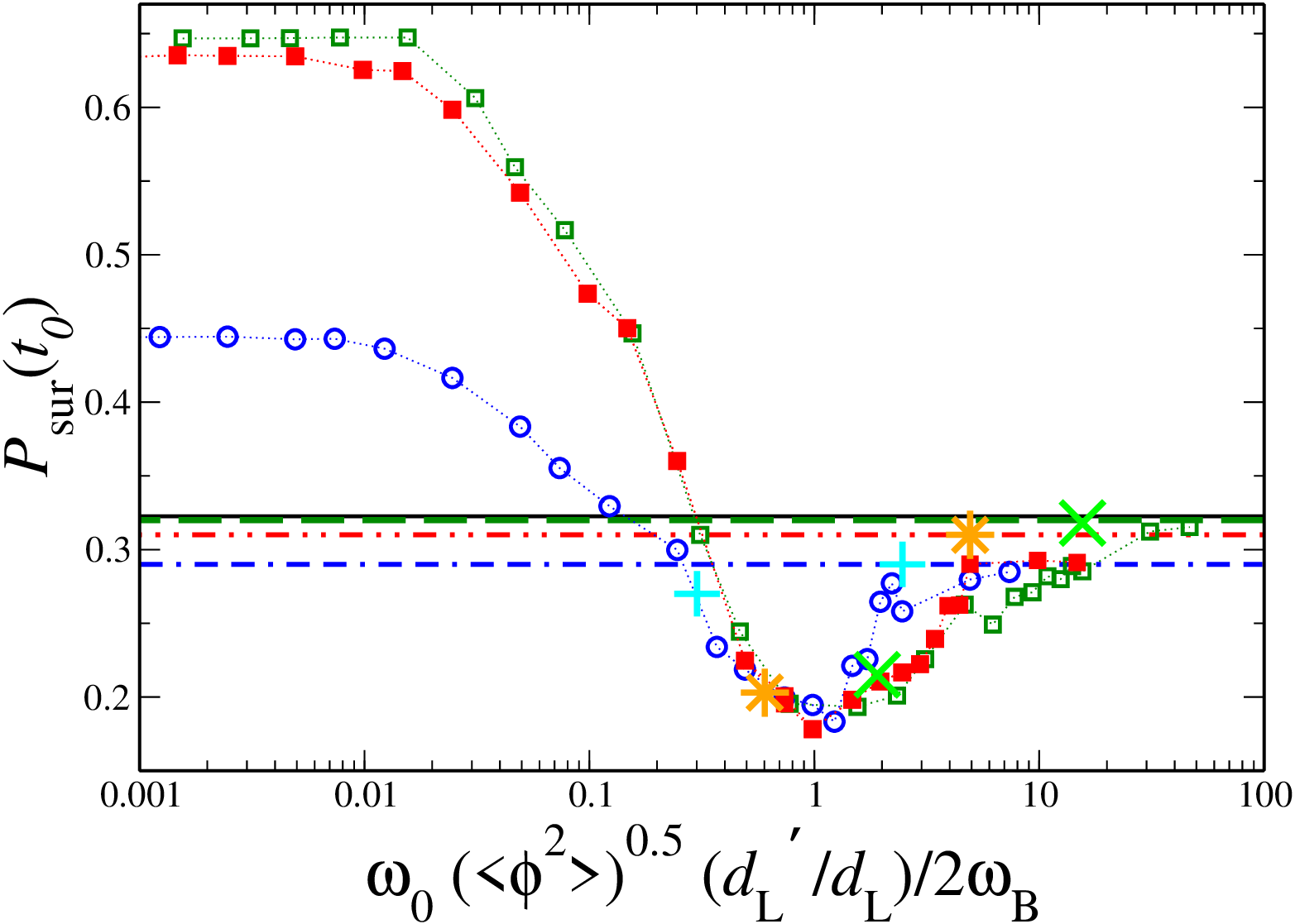}
   \caption{
   (color online) $P_{\rm sur}(t_0\approx 6\;T_{\rm B})$ for $V_{0}=2.5$, $F_{\rm 0}=1.5$, $\Gamma=5/T_{\rm B}$, for fixed noise variances % $\langle\phi^2\rangle=0.01$ (corresponding to $\alpha\approx0.71$, filled circles), 
 $\langle\phi^2\rangle=0.25$ (corresponding to $\alpha\approx0.85$, open circles), $1$ ($\alpha\approx0.99$, filled squares), and $10$ ($\alpha\approx1$, open squares). For comparison we show $P_{\rm sur}(t_0)$ for the respective effective models by the dash-dash-dotted, dot-dot-dashed, and dashed lines (in the same order and color of the symbols), for the reference system (thick solid line), and calculated using a deterministically oscillating phase with $A_{\rm d}^2/2\equiv T/\omega_0^2=0.25$ (pluses), $1$ (stars) and $10$ (crosses).
 } 
   \label{P_surt0_omega_variace}
\end{center}
\end{figure}   

We find that, for a broad range of noise parameters, our main observable, the survival probability $P_{\rm sur}(t_0)$ at fixed time $t_0$, obeys a scaling relation as a function of {\em all} parameters. Our results are similar for other choices of $t_0 \gtrsim 4 T_{\rm B}$, and we stick in the following to $t_0 \approx 6 T_{\rm B}$, a typical experimental observation time reported, e.g., in \cite{Morsch06}. In Fig.~\ref{P_surt0_omega_variace}, the noise frequency $\omega_{0}$ is varied by keeping the variance of the noise fixed. Rescaling now $\omega_{0}$ by $\omega_{\rm B}$, by the standard deviation of the noise $\sqrt{\langle\phi^2\rangle}$ and by the ratio of the two lattice periods $d^{'}_{\rm L}/d_{\rm L}$, all the results of Fig.~\ref{P_surt0_omega_variace} show a similar behavior with a dip at $\omega_0\sqrt{\langle\phi^2\rangle} \approx \,\omega_{\rm B}d^{'}_{\rm L}/(2d_{\rm L})$. First this value of energy lies just above the average band gap, i.e., $ \omega_0\sqrt{\langle\phi^2\rangle} \gtrsim\Delta E\approx 2.5\,E_{\rm rec}$, which turned out to be a necessary condition to observe the dip. Secondly, it implies that the optimal time scale of the noise is directly related to $T_{\rm B}$ for observing an enhancement of the decay 
%(i.e., a dip in  Fig.~\ref{P_surt0_omega_variace}).  
%Intuitively, we expect such a dip for noise histories which facilitate the motion of the wave packet from the center of the Brillouin zone to the band edge. Once the wave packet has reached the band edge, it will easily fall into the next minimum along the band even in the absence of further activation; for a similar argument in noise-driven systems see, e.g., \cite{LM1997}. Dynamically speaking, $T_{\rm B}$ is the time taken in the reference lattice to reach the band edge. Hence the noise is most effective when its typical time scales are of the same order as $T_{\rm B}$. 

The additional factor $d^{'}_{\rm L}/d_{\rm L}$ takes account of the different periods of the two lattices, and the scaling was checked for various incommensurable ratios $d^{'}_{\rm L}/d_{\rm L}$ (not shown). Faster noise is less efficient again since it leads to the regime where the effective model applies, see Eq.~\eqref{eq:eff}. Here the noise effectively averages during the evolution. Indeed at very large $\omega_0$, $P_{\rm sur}(t_0)$ lies on top of the corresponding value (shown by horizontal lines) calculated using the effective potential. The observed scaling relation around the dip, as our main result, drastically reduces the parameter dependence of the system. It helps predict and control the behavior of the survival probability by a simple choice of right combinations of the {\em a priori} many parameters ($F_0, V_0, \omega_0, \Gamma, T$). 

Deviations from the scaling are expected when the noise frequency $\omega_{0}$ becomes comparable with its spectral width, which is determined by $\Gamma$. This implies a crossover from the spectral distribution shown in Fig.~\ref{HN-POT}(b) to the case of Fig.~\ref{HN-POT}(a). Then the noisy phase changes slowly in time, and for $T \lesssim 10/T_{\rm B}^2$ cannot assist the tunneling anymore, as observed in Fig.~\ref{HN-POT}(a). Here $\omega_0\sqrt{\langle\phi^2\rangle} \ll \omega_{\rm B}$ and the interband tunneling depends very much on the strength $T$ of the noise, cf. Fig.~\ref{P_surt0_omega_variace}, since single noise events can dominate the evolution. To better understand the regime where the scaling breaks down, it is helpful to study more explicitly the dependence on $\langle\phi^2\rangle$, which is done in Fig.~\ref{P_surt0_omega_T}. We distinguish two cases: in (a) the data is produced as previously described by adapting $\alpha$ to compare with the reference system ($\alpha$ is calculated as a function of $\beta=\exp(-2\langle\phi^2\rangle)$); in (b) we keep $\alpha$ fixed to understand the effect of the absolute value of the lattice depth (loosing, of course, the meaning of a reference system). The data collapses onto a single curve described by the effective fast noise model of Eq.~\eqref{eq:eff} (dashed lines in Fig.~\ref{P_surt0_omega_T}) for large frequencies $\omega_0^2/T \gtrsim 100$. The behavior of the various curves in (a) is similar (approximate scaling) for $1 < \omega_0^2/T < 100$, provided that $T \lesssim 1/T_{\rm B}^2$. As shown also in Fig.~\ref{HN-POT}(a), for a large $T \geq 100/T_{\rm B}^2$  even the slowly varying noise is strong enough to enhance the decay below the value predicted for the static reference system. For the latter case, the scaling relation obviously is bound to fail. In Fig.~\ref{P_surt0_omega_T}(b), where we do not correct for the change in the relative height of the lattice amplitudes, the just described trends in (a) are shifted and the scaling is harder to appreciate (except for the very fast noise again, where our data always follows the expectation of the effective static model).
\begin{figure}%[H]
\begin{center}
   \includegraphics[width=\linewidth,angle=0]{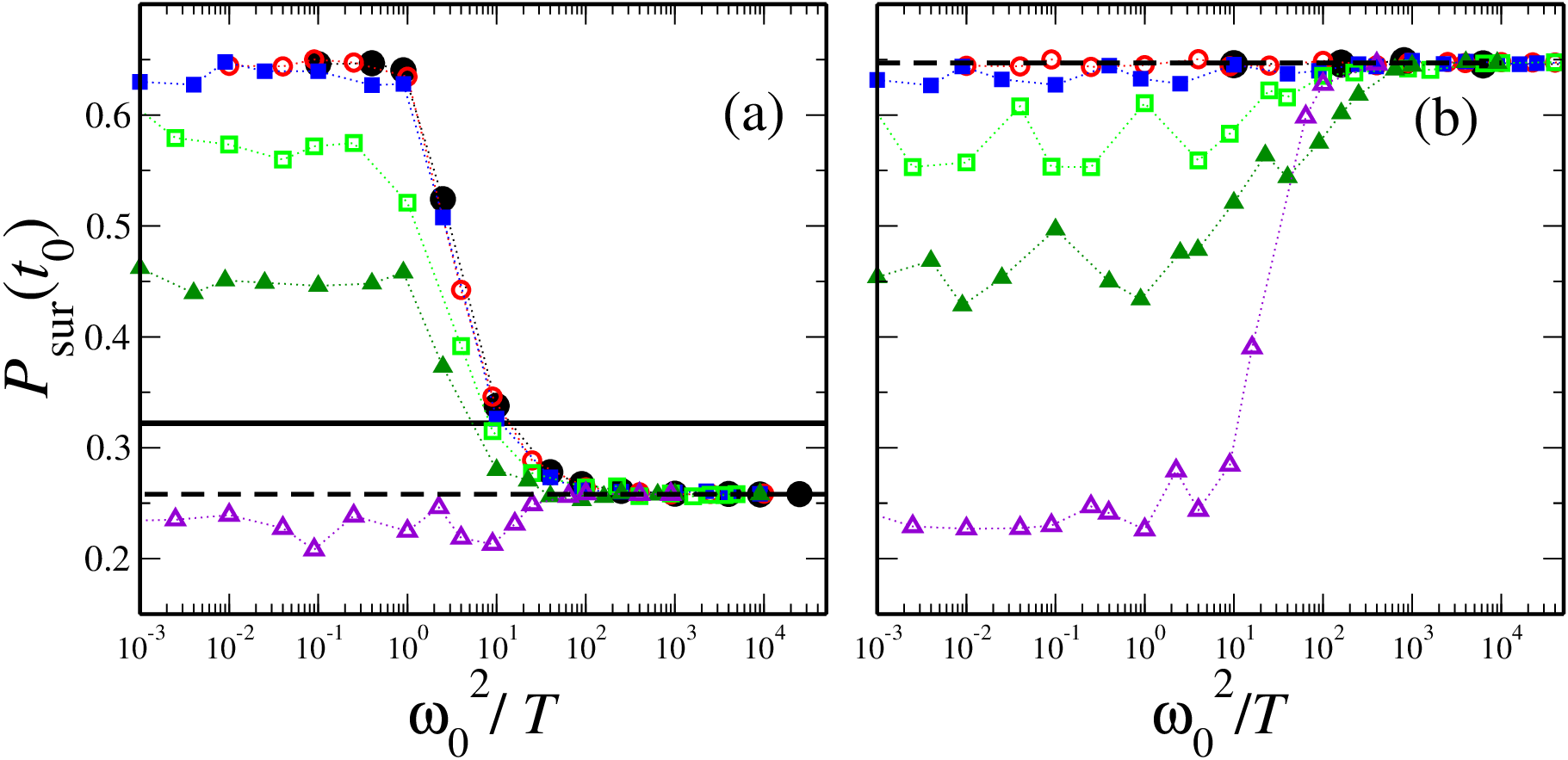}
    \caption{ \small \label{P_surt0_omega_T} (color online) $P_{\rm sur}(t_0\approx6\;T_{\rm B})$ for $V_{0}=2.5$, $F_{0}= 1.5$ vs. $\langle\phi^2\rangle^{-1}=\omega_0^2/T$: (a) adapted $\alpha \neq 1$; (b) fixed $\alpha=1$. Other parameters: $\Gamma=5/T_{\rm B}$, and $T=0.001/T_{\rm B}^2$ (filled circles), $T=0.01/T_{\rm B}^2$ (open circles), $T=0.1/T_{\rm B}^2$ (filled squares), $T=1/T_{\rm B}^2$ (open squares), $T=10/T_{\rm B}^2$ (filled triangles) and $T=100/T_{\rm B}^2$ (open triangles). $P_{\rm sur}(t_0)$ of the corresponding static reference system is shown by the solid line in (a) and for the effective model with $\beta=1$ by the dashed lines in (a,b).} 
\end{center}
\end{figure}

As a final benchmark, we compare to a periodically oscillatory phase $\phi(t)=A_{\rm d}\cos(\omega_{\rm d} t + \varphi)$, with $A_{\rm d}$ being its amplitude and $\omega_{\rm d}$ its oscillation frequency. To model random noise fluctuations the data are averaged over the parameter $\varphi$ (picked randomly from a flat distribution). The spectral distribution of this function is given by $S_{\rm d}\left(\omega\right) = \frac{A^2}{4\pi}\left\lbrace\delta(\omega-\omega_{\rm d})+\delta(\omega+\omega_{\rm d})\right\rbrace$. Practically, the delta functions gain a finite width $\Delta\omega_{min}=\frac{2\pi}{T_{\rm tot}}$, where $T_{\rm tot}$ is the total numerical integration time. The spectral distribution of the harmonic noise in the fast noise regime, which very much peaks at ${\tilde\omega}_0$, is then comparable to the delta-like spectral distribution around $\omega_{\rm d}$ of the oscillating noise. Therefore, for $\omega_{\rm d}\approx\tilde{\omega}_0 \gg \Gamma$ and $A_{\rm d}=\sqrt{2T/\omega_0^2}$, the periodic phase (averaged over $\varphi$) has a similar effect as harmonic noise. Examples of results are shown in Fig.~\ref{P_surt0_omega_variace} for $A_{\rm d}^2/2=0.25$, $1$ and $10$, respectively. As expected, for large $\omega_0$, the symbols lie again on top of the prediction of the effective model.

We investigated the impact of noise on the Landau--Zener tunneling in a one-dimensional Wannier-Stark system. 
%We motivated and derived an effective model (cf., Eq.~\eqref{eq:eff}) which can be used as a benchmark to describe our results in the regime of fast oscillating noise. 
By a proper scaling of the data with parameters, we can universally characterize the effect of the noise. This central result is summarized in Fig.~\ref{P_surt0_omega_variace}. Our predictions show that time-dependent noise in a bichromatic lattice provides a further handle to control the transport from the ground band to higher energy bands. We describe a first step in the direction to push investigations of the static regime into the realm of dynamical disorder \cite{Timenoise}. A natural extension would be to study the case of the simultaneous presence of noise and atom-atom interactions in our Wannier-Stark system, complementary to expansion experiments in flat non-tilted potentials in optics \cite{Moti2011} and with BEC \cite{Lucioni10}.

We are very grateful to E. Arimondo and O. Morsch for inspiring discussions and for support by the
Excellence Initiative through the HGSFP (grant number GSC 129/1), the Frontier Innovation Fund, the Global Networks Mobility Measures, and the DFG FOR760.

\bibliographystyle{apsrmp}

\end{document}